\documentclass[12pt,a4paper,epsf]{article}
\usepackage{graphics}
\usepackage{amssymb,amsmath}
\usepackage[dvips]{lscape,graphicx}
\usepackage{cite}

\newcommand{\ct}{\cite}
\newcommand{\lb}{\label}

\newcommand{\bc}{\begin{center}}
\newcommand{\ec}{\end{center}}
\newcommand{\bd}{\begin{displaymath}}
\newcommand{\ed}{\end{displaymath}}
\newcommand{\be}{\begin{equation}}
\newcommand{\ee}{\end{equation}}
\newcommand{\ba}{\begin{array}}
\newcommand{\ea}{\end{array}}
\newcommand{\bea}{\begin{eqnarray}}
\newcommand{\eea}{\end{eqnarray}}
\newcommand{\bt}{\begin{tabular}}
\newcommand{\et}{\end{tabular}}

\newcommand{\bp}{\begin{picture}}
\newcommand{\ep}{\end{picture}}
\newcommand{\bfi}{\begin{figure}}
\newcommand{\efi}{\end{figure}}
\vspace{0.5cm}

\def\fun#1#2{\lower3.6pt\vbox{\baselineskip0pt\lineskip.9pt
\ialign{$\mathsurround=0pt#1\hfil##\hfil$\crcr#2\crcr\sim\crcr}}}

\begin{document}

\hyphenation{ }

\title{\huge \bf {Generalized Dual Symmetry of Nonabelian
Theories,\\
Monopoles and Dyons\\[5mm]}}

\author{\LARGE \bf C.R.~Das\\[3mm]
\Large The Institute of Mathematical Sciences,\\[3mm]
\Large Chennai, India\\[5mm]
\LARGE \bf L.V.~Laperashvili\\[3mm]
\Large The Institute of Theoretical and Experimental Physics,\\[3mm]
\Large Moscow, Russia\\[5mm]
\LARGE \bf H.B.~Nielsen \\[3mm]
\Large The Niels Bohr Institute,\\[3mm]
\Large Copenhagen, Denmark\\[25mm]
\large \bf A talk presented at the\\[3mm]
\LARGE \bf 12th Lomonosov Conference\\[3mm]
\LARGE \bf on Elementary Particle Physics,\\[3mm]
\large \bf dedicated to 250 years jubilee of the Moscow State
University,\\[3mm]
\large \bf Moscow, 25--31 August, 2005.}

\date{}

\maketitle

\thispagestyle{empty}

\clearpage\newpage

\thispagestyle{empty}

\begin{abstract}

In the present talk we present an investigation of nonabelian $SU(N)$ gauge theories, 
describing a system of fields with non--dual $g$ and dual $\tilde g$ charges and revealing the 
generalized dual symmetry. The Zwanziger type action is suggested. The renormalization group 
equations for pure nonabelian theories, in particular for pure $SU(3)\times\widetilde{SU(3)}$ 
gauge theory (as an example) are analysed. We consider not only monopoles, but also dyons. The 
behaviour of the QCD total beta--function is investigated. It was shown that this beta--function
is antisymmetric under the interchange $\alpha\leftrightarrow\frac 1\alpha$ (here 
$\alpha\equiv\alpha_s$), and has zero (``fixed point'') at $\alpha = 1$. Monopoles, or dyons, 
are responsible for the phase transition. Considering critical points at $\alpha_1\approx 0.4$ 
and $\alpha_2\approx 2.5$, we give an explanation of the freezing of $\alpha_s$.

\end{abstract}

\clearpage\newpage

\parindent=1cm

\thispagestyle{empty}

\section*{Contents}

\vspace{3cm}

{ \bf
\begin{itemize}
\item[1.] Introduction: Loop space variables of\\ nonabelian theories.

\item[2.] The Zwanziger--type action of nonabelian
\\ theories and duality.

\item[3.] The charge quantization conditions.

\item[4.] Renormalization group equations and duality.

\item[5.] An example of beta--function in the case of
the\\ pure $SU(3)$ colour gauge group (Part I).

\item[6.] The ``abelization" of monopole vacuum of\\
nonabelian gauge theories.

\item[7.] An example of beta--function in the case of
the\\ pure $SU(3)$ colour gauge group (Part II).

\item[8.] Dyons.

\item[9.] Conclusions.
\end{itemize}}

\clearpage\newpage
\pagenumbering{arabic}

\section{Introduction: Loop space variables of nonabelian theories}

In the last years gauge theories essentially operate with the fundamental idea of duality.

In the present investigation we consider the $SU(N)$ nonabelian theories in terms of loop 
variables. For the standard (non--dual) sector we consider the path ordered exponentials:
$$\Phi(C)=P\exp\left[ ig\oint A_{\mu}(\xi)d{\xi}^{\mu}\right]=P\exp\left[ ig\int_0^{2\pi}
A_{\mu}(\xi)\dot{\xi}^{\mu}(s)ds\right],$$
\begin{figure}[h]
\centering
\includegraphics[height=90mm,keepaspectratio=true]{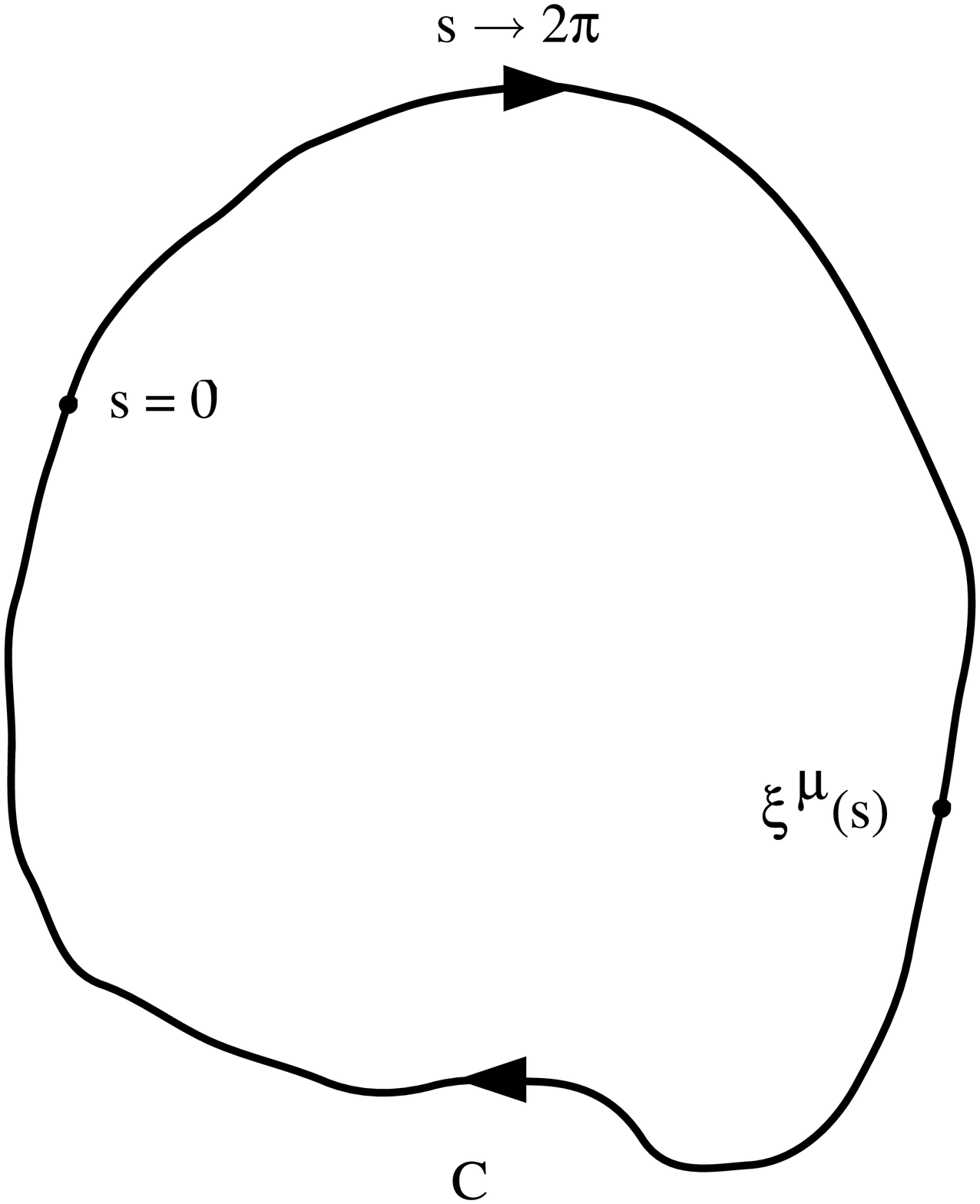}
\caption{}
\end{figure}
where $C$ is a parameterized closed loop with coordinates ${\xi}^{\mu}(s)$ in 4--dimensional 
space (see Figure 1). The loop parameter is $s$: $0\le s \le 2\pi$;
$$\dot {\xi}^{\mu}(s)=\frac {d{\xi}^{\mu}(s)}{ds}.$$
We also consider the following unclosed loop variables:
$$\Phi(s_1,s_2)=P\exp\left[ ig\int_{s_1}^{s_2}A_{\mu}(\xi)\dot{\xi}^{\mu}(s)ds\right].$$
Therefore
$$\Phi(C)\equiv\Phi(0,2\pi).$$

For the dual sector we have:
$$\tilde\Phi(\tilde C)=P\exp\left[ i\tilde g\oint{\tilde A}_{\mu}(\eta)d{\eta}^{\mu}\right]
=P\exp\left[ i\tilde g\int_0^{2\pi}{\tilde A}_{\mu}(\eta)\dot{\eta}^{\mu}(t)dt\right],$$
where $\tilde C$ is a parameterized closed loop in the dual sector with coordinates 
${\eta}^{\mu}(t)$ in the 4--dimensional space (see Figure 2).
\begin{figure}[h]
\centering
\includegraphics[height=90mm,keepaspectratio=true]{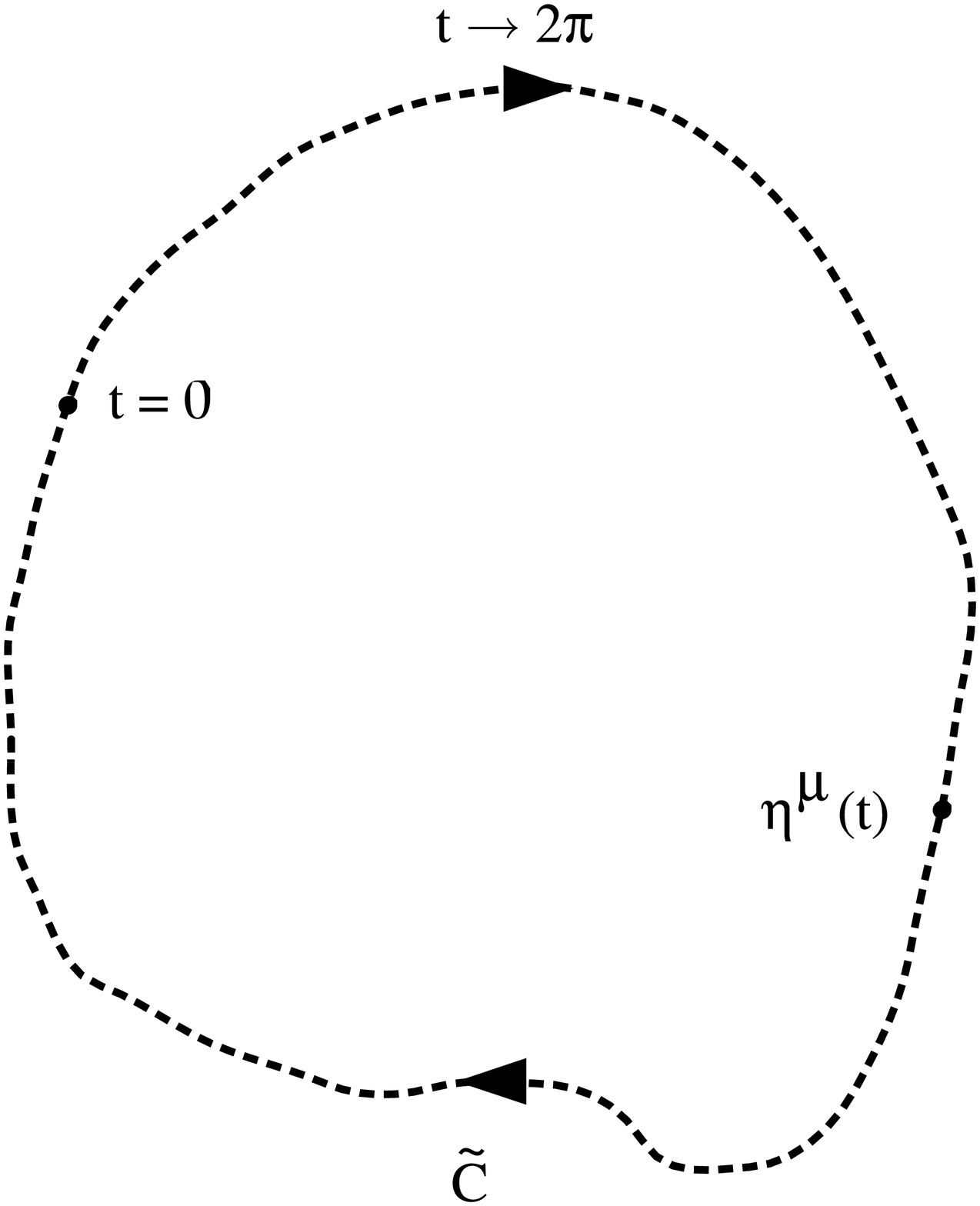}
\caption{}
\end{figure}
The loop parameter is $t$: $0\le t \le 2\pi$;
$$\dot {\eta}^{\mu}(t)=\frac {d{\eta}^{\mu}(t)}{dt}.$$
The unclosed loop variables in the dual sector are:
$$\tilde\Phi(t_1,t_2)=P\exp\left[ i\tilde g\int_{t_1}^{t_2}{\tilde A}_{\mu}
(\eta)\dot{\eta}^{\mu}(t) dt\right].$$
Therefore
$$\tilde\Phi(\tilde C)\equiv\tilde\Phi(0,2\pi).$$
Here standard and dual sectors have coupling constants $g$ and $\tilde g$, respectively.

Considering, for simplicity of presentation, only gauge groups $SU(N)$, we have 
vector--potentials $A_{\mu}$ and $\tilde A_{\mu}$ belonging to the adjoint representation of 
the $SU(N)$ and $\widetilde{SU(N)}$ groups:
$$A_{\mu}(x)=A_{\mu}^j t^j, \qquad\tilde A_{\mu}(x)=\tilde A_{\mu}^j t^j, 
\qquad j=1,...,N^2 - 1,$$
where $t^j$ are generators of the $SU(N)$ group. As a result, we consider nonabelian theories 
having a doubling of symmetry from $SU(N)$ to
$$SU(N)\times\widetilde{SU(N)}.$$

\section{The nonabelian Zwanziger--type action and duality}

Following the idea of Zwanziger \ct{1} to describe symmetrically non--dual and dual abelian 
fields $A_{\mu}$ and $\tilde A_{\mu}$, covariantly interacting with electric $j_{\mu}^{(e)}$ 
and magnetic currents $j_{\mu}^{(m)}$, respectively, we suggest to construct the generalized 
Zwanziger formalism for the pure nonabelian gauge theories, considering the following
Zwanziger--type action:
\bea S &=& - \frac 2K \int {\cal D}{\xi}^{\mu}ds \left\{ Tr\left(E^{\mu}[\xi|s] E_{\mu}[\xi|s]
\right) + Tr \left({\tilde E}^{\mu}[\xi|s]{\tilde E}_{\mu}[\xi|s]\right)\right.\nonumber\\
&&+\left. i Tr\left(E^{\mu}[\xi|s]{\tilde E}_{\mu}^{(d)}[\xi|s]\right) + i Tr\left({\tilde E}
^{\mu}[\xi|s]E_{\mu}^{(d)}[\xi|s]\right)\right\} {\dot\xi}^{-2}(s) + S_{gf},\lb{1}\eea 
Here we have used the Chan--Tsou variables \ct{2}:
\be E_{\mu}[\xi|s]=\Phi(s,0)F_{\mu}[\xi|s]{\Phi}^{-1}(s,0),\lb{2}\ee 
where 
\be F_{\mu}[\xi|s]=\frac{i}{g}{\Phi}^{-1}(C(\xi))\frac{\delta\Phi(C(\xi))}
{\delta\xi^{\mu}(s)}\lb{3}\ee 
are the Polyakov variables. Using $\tilde\Phi$, we have the analogous expressions for 
${\tilde F}_{\mu}[\xi|s]$ and ${\tilde E}_{\mu}[\xi|s]$. In Eq.~(2) $K$ is the normalization 
constant: 
\be K=\int_0^{2\pi}ds{\Pi}_{s'\neq s}d^4\xi(s'),\lb{4}\ee $S_{gf}$ is the gauge--fixing action, 
excluding ghosts in theory:
\be S_{gf}=\frac 2K\int{\cal D}\xi^{\mu}ds\left[ M_A^2 {\left(\dot{\xi}\cdot A\right)}^2 +
M_B^2{\left({\dot\xi}\cdot\tilde A\right)}^2\right]{\dot\xi}^{-2}.\lb{5}\ee 

Also we have used the Chan--Tsou generalized dual operation \ct{2}:
$$E_{\mu}^{(d)}[\xi|s]=$$
\be -\frac 2K\epsilon_{\mu\nu\rho\sigma}{\dot\xi}^{\nu}\int{\cal D}{\eta}^{\mu}dt\omega 
(\eta(t))E^{\rho}[\eta|t]{\omega}^{-1}(\eta(t)){\dot\eta}^{\sigma}(t){\dot\eta}^{-2}
 \delta (\eta(t) - \xi(s)).\lb{6}\ee 
The last integral in Eq.~(6) is over all loops and over all points of each loop, and the factor 
$\omega(x)$ is just a rotational matrix allowing for the change of local frames between the two 
sets of variables. In the abelian case expression (6) coincides with the Hodge star operation:
$$F_{\mu\nu}^{*} = \frac 12\epsilon_{\mu\nu\rho\sigma}F_{\rho\sigma},$$
but for nonabelian theories they are different.

From our Zwanziger--type action we have the following equations of motions: 
\be\frac{\delta E_{\mu}[\xi|s]}{\delta \xi^{\mu}(s)}=0,\qquad
\frac{\delta\tilde E_{\mu}[\xi|s]}{\delta\xi^{\mu}(s)}=0.\lb{7}\ee 
Such a theory shows the invariance under the generalized dual operation, that is, has a dual 
symmetry under the interchange: 
\be E_{\mu}\longleftrightarrow{\tilde E}_{\mu}.\lb{8}\ee 
Here ${\tilde E}_{\mu}$ is given by the Chan--Tsou dual operation: 
\be {\tilde E}_{\mu}=E^{(d)}_{\mu}.\lb{9}\ee

\section{The charge quantization condition}

Considering the Wilson operator:
$$ A(C)=Tr\left(P\exp \left[ig\oint_C A_{\mu}(\xi) d{\xi}^{\mu}\right]\right),$$
which measures chromo--magnetic flux through $C$ and creates chromo--electric flux along $C$,
also considering the dual operator:
$$B(\tilde C)=Tr\left(P \exp \left[ i\tilde g\oint_C {\tilde 
A}_{\mu}(\eta)d{\eta}^{\mu}\right]\right),$$
which measures chromo--electric flux through $\tilde C$ and has chromo--magnetic flux along 
$\tilde C$, we can use (following Ref.~\ct{2}) the t'Hooft commutation relation: 
\be A(C)B(\tilde C)=B(\tilde C)A(C)\exp(2\pi n/N),\ee 
where $n$ is the number of times $\tilde C$ winds around $C$ and $N\ge 2$ is for the gauge 
group $SU(N)$. This t'Hooft relation produces the generalized condition for the charge 
quantization: 
\be g\tilde g=4\pi n,\qquad n\in Z,\ee
so called the Dirac--Schwinger--Zwanziger (DSZ) relation.

Using constants containing the elementary charges $g$ and $\tilde g$ (the case $n=1$): 
\be \alpha=\frac{g^2}{4\pi},\qquad\tilde\alpha=\frac{{\tilde g}^2}{4\pi},\lb{11}\ee 
we have the following relation:
\be\alpha\tilde\alpha=1.\lb{12}\ee

\section{Renormalization group equations and duality}

For pure nonabelian gauge theories, duality gives a symmetry under the interchange: 
\be\alpha\leftrightarrow\tilde\alpha,\qquad{\mbox{or}}\qquad\alpha\leftrightarrow 
\frac{1}{\alpha},\lb{37}\ee 
what follows from the relation (\ref{12}). For the first time such a symmetry was considered 
in the Yang--Mills theories by Montonen and Olive \ct{3}.

In nonabelian theories with chromo--electric and chromo--magnetic charges the derivatives 
$$d\ln\alpha/dt\qquad {\mbox{and}}\qquad d\ln\tilde\alpha/dt$$ 
are only functions of the effective constants $\alpha(t)$ and $\tilde\alpha(t)$, as in the 
Gell--Mann--Low theory. Here $t=\ln (\mu^2/M^2),$ $\mu$ is the energy variable and $M$ is the 
renormalization scale.

In general case, we can write the following RGEs \ct{4,4a}: 
\be\frac{\mbox{d}\ln\alpha(t)}{\mbox{dt}}=-\frac{\mbox{d}\ln {\tilde\alpha}(t)}{\mbox{dt}}
=\beta(\alpha )-\beta(\tilde\alpha)=\beta^{(total)}(\alpha),\lb{41}\ee
which comes from the dual symmetry and charge quantization condition, valid for arbitrary $t$: 
$\alpha(t)\tilde\alpha(t)=1$. In Eq.~(\ref{41}) $\beta(\alpha)$ is the well--known perturbative
beta--function.

Here we see that the total beta--function for the pure nonabelian theory:
\be\beta^{(total)}(\alpha)=\beta(\alpha) - \beta(\tilde\alpha)\lb{42}\ee
is antisymmetric under the interchange (\ref{37}), what means that $\beta^{(total)}(\alpha)$ 
has zero (``fixed point") at $\alpha = \tilde \alpha = 1$:
\be\beta^{(total)}(\alpha=\tilde\alpha = 1)=0.\lb{43}\ee

\section{An example of beta--function in the case of the pure $SU(3)$ colour gauge group 
(Part I)}

The investigation of gluondynamics -- the pure $SU(3)$ colour gauge theory -- shows that at 
sufficiently small $\alpha < 1$, the $\beta$--function in the 3--loop approximation is given
by the following series over $\alpha /4\pi$ \ct{5}:
\be\beta(\alpha )= - \left[\beta_0\frac{\alpha}{4\pi} + \beta_1{\left(\frac{\alpha}{4\pi}
\right)}^2 + \beta_2{\left(\frac{\alpha}{4\pi}\right)}^3 + ...\right], \lb{43}\ee 
where
\be\beta_0 = 11,\qquad\beta_1 = 102,\qquad\beta_2 = 1428.5.\lb{44}\ee 
QCD shows that $\alpha_s$ is freezing at the value $\alpha_s\approx 0.4$ \ct{6}.
Assuming the following freezing QCD coupling constants:
$$\beta(\alpha) = 0 \qquad {\mbox{for}}\qquad\alpha > 0.4,\qquad {\mbox{and}}\qquad 
\beta(\tilde \alpha) = 0 \qquad {\mbox{for}}\qquad\tilde\alpha > 0.4,\qquad {\mbox{or}}$$ 
$$\beta(\alpha) = 0 \qquad {\mbox{for}}\qquad 0.4 < \alpha,\tilde\alpha < \frac 1{0.4} = 2.5,$$
we have the behaviour of $\beta^{(total)}(\alpha)= \beta(\alpha) - \beta(\tilde \alpha) 
= \beta(\alpha) - \beta(1/\alpha)$, given by Figure 3.
\begin{figure}[h]
\centering
\includegraphics[height=150mm,keepaspectratio=true,angle=-90]{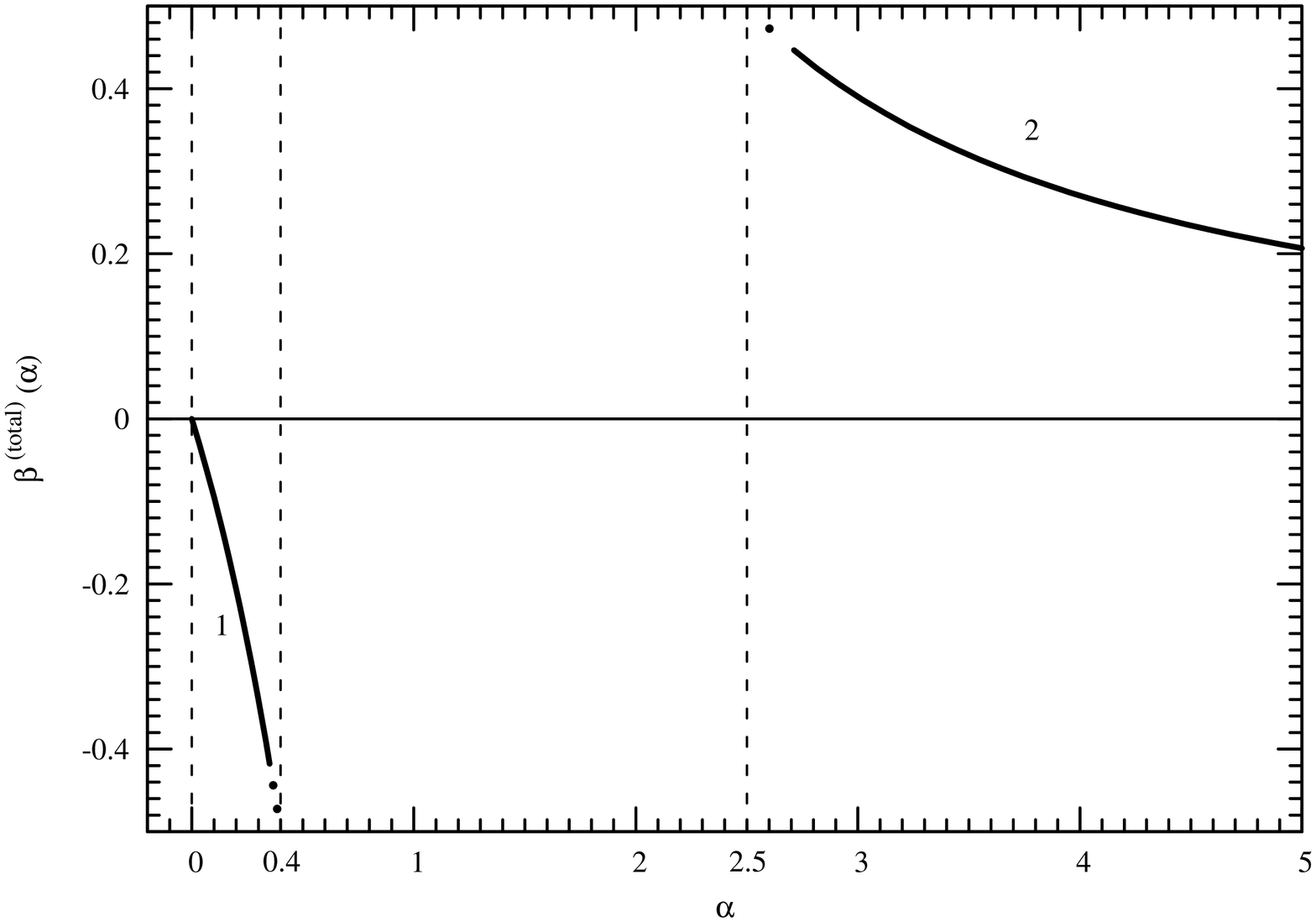}
\caption{}
\end{figure}

\section{The ``abelization" of monopole vacuum of nonabelian gauge theories}

In the light of contemporary ideas of the abelization of the $SU(N)$ gauge theories \ct{7} 
(see also \ct{8}), it is attractive to consider the behaviour of $\beta^{(total)}(\alpha)$ in 
the vicinity of the point $\alpha = 1$.

Lattice investigations of the pure $SU(3)$ theories show that in some region 
$\alpha > \alpha_{conf}$ non--perturbative effects lead to the phenomenon when the gauge fields 
$A_{\mu}^J$ (here $J=1,...8$) make up composite configurations of monopoles, which form a 
monopole condensate. Such a condensate creates strings between chromo--electric charges, which 
confine these charges. It is natural to think that the same configurations are created in the 
local $SU(3)$ gauge theory and imagine them as the Higgs fields $\tilde \phi (x)$ of scalar
chromo--magnetic monopoles. T'Hooft \ct{7} suggested to consider such a gauge, in which 
monopole degrees of freedom, hidden in composite monopole configurations, become explicit and 
abelian.

He developed {\it a method of the Maximal Abelian Projections (MAP).} According to this method, 
in the non--perturbative region scalar monopoles are created only by the diagonal $SU(3)$
components ${({ A}_{\mu})}^i_i$ of gauge fields ${({A}_{\mu})}^i_j$, where $i,j = 1,2,3$ are 
colour indices, and interact only with the diagonal $\widetilde{SU(3)}$ components of gauge 
fields ${({\tilde A}_{\mu})}^i_j$.

In the non--perturbative region, the non--diagonal $SU(3)$ and $\widetilde{SU(3)}$ components 
of gauge fields are suppressed and the interaction of monopoles with dual gluons is described by
$\widetilde {U(1)}\otimes \widetilde {U(1)}$ (Cartain) subgroup of $\widetilde {SU(3)}$ group.
These monopoles can be approximately described by the Higgs fields $\tilde \phi(x)$ of scalar 
chromo--magnetic monopoles, interacting with gauge fields $\tilde A_{\mu}$.

Recalling the generalized dual symmetry, we are forced to assume that the similar composite 
configurations have to be produced by dual gauge fields $\tilde A_{\mu}^J$, and described by the
Higgs fields $\phi(x)$ of scalar chromo--electric ``monopoles", interacting with gauge fields 
$A_{\mu}$. The interaction of ``monopoles" with gluons also has to be described by 
$U(1)\otimes U(1)$ subgroup of $SU(3)$ group.

In general, $U(1)$--subgroups are arbitrary embedded into the $SU(N)$ gauge group:
$U(1)^{N-1}\subset SU(N)$, and in the non--perturbative region $SU(N)$ gauge theory is reduced 
to the abelian $U(1)^{N-1}$ theory with $N-1$ different types of abelian monopoles.

The generators of the Cartain subgroup are given by the diagonal
Gell--Mann matrices:
$$ t^3 = \frac{\lambda^3}{2}\qquad {\mbox{and}}\qquad t^8 = \frac{\lambda^8}{2}.$$
Thus, in the non--perturbative region we have the following equations for the diagonal 
$F_{\mu\nu},\phi$ and $\tilde \phi$ \ct{4a}: 
\be\partial_{\nu}F_{\mu\nu}^{J=3,8} =\frac{i}{2} g\left[\phi^{+}\left(\frac{\lambda^{3,8}}
{2}\right){\cal D}_{\mu}\phi - (\cal D_{\mu}\phi)^+\left(\frac{\lambda^{3,8}}{2}\right)
\phi\right],\lb{7h}\ee 
and 
\be\partial_{\nu}{\tilde F}_{\mu\nu}^{J=3,8} = \frac{i}{2}\tilde g\left[\tilde \phi^{+}\left
(\frac{\lambda^{3,8}}{2}\right){\tilde {\cal D}}_{\mu}\tilde\phi - ({\tilde {\cal D}}_{\mu}
\tilde\phi)^{+}
\left(\frac{\lambda^{3,8}}{2}\right)\tilde\phi\right],\lb{8h}
\ee 
where $$ {\cal D}_{\mu} = \partial_{\mu} - igA_{\mu}\quad {\mbox{and}}\quad
      {\tilde {\cal D}}_{\mu} = \partial_{\mu} - i{\tilde g}{\tilde A}_{\mu}.  $$
Here we can choose two independent abelian monopoles as:
$${\tilde \phi}_1 = ({\tilde\phi})^1_1\qquad {\mbox{and}}\qquad {\tilde\phi}_2 = 
({\tilde \phi})^2_2,$$ 
and also two independent abelian scalar fields with electric charges:
$${\phi}_1 = ({\phi})^1_1\qquad {\mbox{and}}\qquad {\phi}_2 = ({\phi})^2_2.$$

Considering the radiative corrections to the gluon propagator shown in Figure 4, it is easy to 
calculate that both, abelian monopoles ${\tilde \phi}_{1,2}$ and ``monopoles" ${\phi}_{1,2}$, 
have the following charges $g_{(MAP)}$ and ${\tilde g}_{(MAP)}$:
\be\alpha_{(MAP)} =\frac{\alpha}{2},\qquad {\tilde\alpha}_{(MAP)}=\frac{\tilde\alpha}{2}.
                                        \lb{9h}\ee
\begin{figure}[h]
\centering
\includegraphics[height=80mm,keepaspectratio=true]{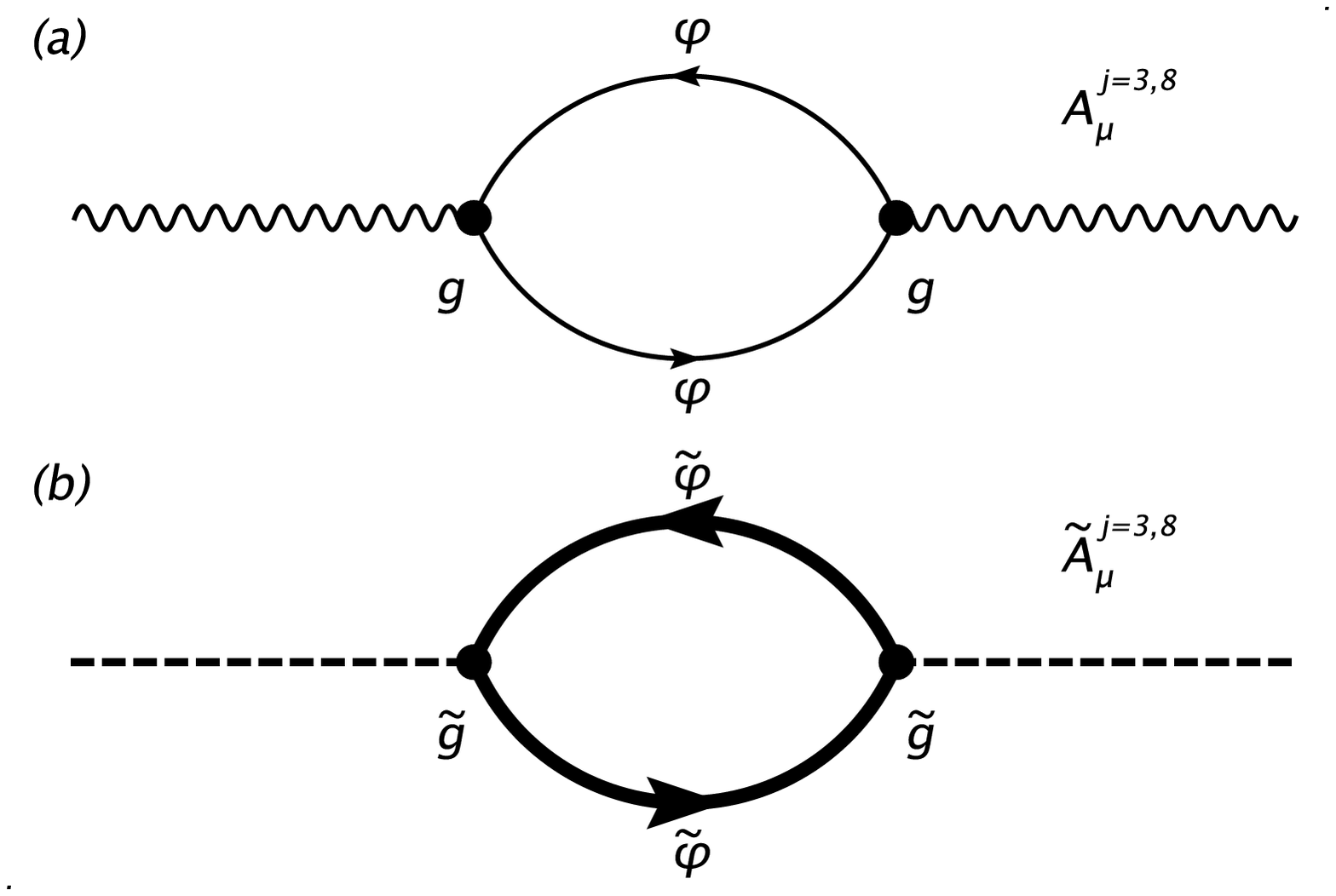}
\caption{}
\end{figure}

Using notations: $f_{\mu\nu,i}\equiv (F_{\mu\nu})^i_i$, $\,a_{\mu,i}\equiv (A_{\mu})^i_i$
and $\,{\tilde a}_{\mu,i}\equiv ({\tilde A}_{\mu})^i_i$,
we have the following equations valid into the non--perturbative region of QCD 
$(i=1,2)$: 
\be\partial_{\nu}f_{\mu\nu,i} =\frac{i}{2} g_{(MAP)}\left[\phi_i^{+}{\cal D}_{\mu}\phi_i -
({\cal D}_{\mu}\phi_i)^{+}\phi_i\right],\lb{10h}\ee
and
\be\partial_{\nu} f^{*}_{\mu\nu,i} = \frac{i}{2}{\tilde g}_{(MAP)}\left[{\tilde\phi}_i^{+}
{\tilde {\cal D}}_{\mu}{\tilde \phi}_i - ({\tilde {\cal D}}_{\mu}{\tilde\phi}_i)^{+}
{\tilde \phi}_i\right], \lb{11h}
\ee where 
$$ {\cal D}_{\mu} = \partial_{\mu} - ig\,a_{\mu},
   \quad  {\tilde {\cal D}}_{\mu} = \partial_{\mu} - i{\tilde g}\,{\tilde a}_{\mu}.   $$

In the case of scalar electrodynamics, which is an Abelian (A) gauge theory, we have the 
following beta--function in the two--loop approximation: 
\be\beta_A(\alpha^{(em)}) = \frac{\alpha^{(em)}}{12\pi}\left(1 + 3\frac{\alpha^{(em)}}
{4\pi} + ...\right).\lb{42}\ee 
For this abelian theory we have the Dirac relation:
$$\alpha^{(em)}{\tilde \alpha}^{(em)} = \frac 14,$$ 
and the following RGEs for electric and magnetic fine structure constants:
$$\frac{\mbox{d}\ln \alpha^{(em)}(t)}{\mbox{d}t} = - \frac{\mbox{d}\ln {\tilde \alpha}^{(em)}
(t)}{\mbox{d}t} = \beta_A(\alpha^{(em)}) - \beta_A({\tilde\alpha}^{(em)})$$
\be = \frac{\alpha^{(em)} - {\tilde \alpha}^{(em)}}{12\pi}\left(1 + 3\frac{\alpha^{(em)} 
+ {\tilde \alpha}^{(em)}}{4\pi} + ...\right).\lb{43}\ee 
As it was shown in Refs.~\ct{4}, the last RGEs can be considered by perturbation theory
simultaneously only in the small region (approximately): 
\be 0.2\stackrel{<}{\sim}\alpha^{(em)}, {\tilde\alpha}^{(em)}\stackrel{<}{\sim }1.\lb{44}\ee
This is valid for all abelian theories.

The behaviour of the effective fine structure constants was investigated in the vicinity of 
the phase transition point in compact lattice QED by the Monte Carlo simulation method \ct{10}.
The following result was obtained:
\be\alpha_{crit}^{lat.QED}\approx 0.20\pm 0.015,\qquad\tilde\alpha_{crit}^{lat.QED}
\approx 1.25\pm 0.10,\lb{45}\ee 
which is very close to the perturbative region (\ref{44}) for parameters $\alpha^{(em)}$ and 
${\tilde \alpha}^{(em)} $.

Using the two--loop approximation for the effective potential in the Higgs model of dual 
scalar electrodynamics, in Ref.~\ct{11} we have obtained the following result:
\be\alpha_{crit}^{(em)}\approx 0.21,\qquad {\tilde \alpha}_{crit}^{(em)}\approx 1.20.\lb{46}\ee 
These values also are very close to the above--mentioned region (\ref{44}). Then our abelian 
monopoles, arising in QCD as a result of MAP, have the following critical constant value: 
\be {\tilde \alpha}_{(MAP)}^{(crit)}\approx 1.25,\lb{47}\ee 
what gives the beginning of the confinement region for QCD:
\be\alpha_1 = \alpha_{conf}= \frac 1{{\tilde\alpha}^{(crit)}} = \frac 1{2{\tilde 
\alpha}_{(MAP)}^{(crit)}}\approx\frac 1{2.5} = 0.4.\lb{48}\ee 
We have received an explanation of the value of freezing $\Large\bf\alpha$.

By dual symmetry, the end of the perturbative region for the scalar field $\phi$ is:
${\tilde \alpha}_{conf} \approx 0.4$, what corresponds to $\alpha_2 = 1/{\alpha_1}\approx 2.5$.

\section{An example of beta--function in the case of the pure $SU(3)$ colour gauge group 
(Part II)}

The last investigation shows that in the region: 
\be 0.4\stackrel{<}{\sim }\alpha, \tilde{\alpha} \stackrel{<}{\sim }2.5\lb{49}\ee 
we have an abelian theory (abelian dominance) with the two scalar monopole fields 
${\tilde \phi}_{1,2}$ and two scalar fields ${\phi}_{1,2}$. The corresponding beta--functions 
are: 
$$\frac{\mbox{d}\ln {\alpha}_{(MAP)}(t)}{\mbox{d}t} = - \frac{\mbox{d}\ln 
{\tilde \alpha}_{(MAP)}(t)}{\mbox{d}t}$$
\be = 2\left[\frac{{\alpha}_{(MAP)} - {\tilde \alpha}_{(MAP)}}{12\pi}\left(1 + 
3\frac{{\alpha}_{(MAP)} + {\tilde\alpha}_{(MAP)}}{4\pi} + ...\right)\right],\lb{50}\ee 
what gives the following beta--functions: 
\be\frac{\mbox{d}\ln\alpha(t)}{\mbox{d}t} = - \frac{\mbox{d}\ln\tilde\alpha (t)}{\mbox{d}t}
 = \frac{\alpha - \tilde \alpha}{12\pi}\left(1 + 3\frac{\alpha + \tilde \alpha}{8\pi} 
+ ...\right),\lb{50}\ee 
valid in the region (\ref{49}). 

The behaviour of the total beta--function for the pure $SU(3)$ colour gauge theory is given by 
curves 1,2,3 in Figure 5 (curve 1' corresponds to QCD), where the regions of the formation of 
strings also are shown.
\begin{figure}[h]
\centering
\includegraphics[height=150mm,keepaspectratio=true,angle=-90]{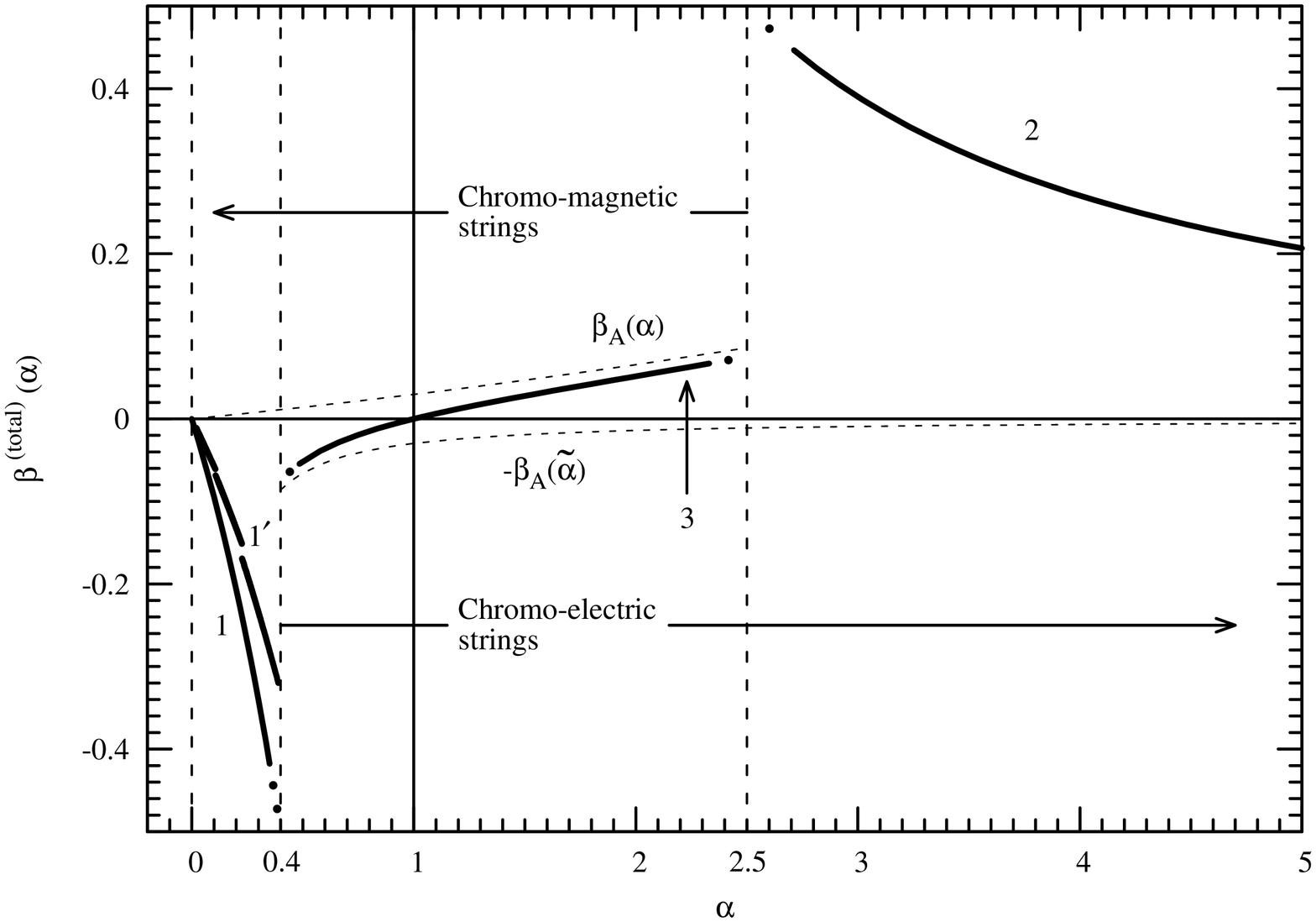}
\caption{}
\end{figure}

Of course, we do not know the behaviour of $\beta$--functions near the phase transition points. 
But this points explain the approximate freezing of $\alpha$ in the region (\ref{49}), where 
both charges, chromo--electric and chromo--magnetic ones, are confined. Chromo--electric 
strings exist for $\alpha > 0.4$ , but chromo--magnetic ones exist for $\alpha < 2.5$, what is 
shown in Figure 5. 

Here we see that the total beta--function has zero at the point $\alpha=\tilde\alpha=1,$ 
predicted by our theory.

\section{Dyons}

The dual symmetry of the pure nonabelian theories leads to the natural assumption that in the 
non--perturbative region, not monopoles and ``monopoles", but dyons are responsible for the
confinement \ct{12}. Then we have united Higgs abelian scalar dyon fields $\phi_{1,2,...N-1}$, 
having simultaneously electric and magnetic charges, and the following equations for each 
components $i=1,2,..., N-1$ can be considered: 
\be\partial_{\lambda} f_{\lambda\mu,i} = ig_{(MAP)}\left[\phi^+_i{\cal D}_{\mu}\phi_i - \phi_i
{\cal D}_{\mu}\phi^+_i\right], \lb{51} \ee
and 
\be\partial_{\lambda} f_{\lambda\mu,i}^{*} = 
i{\tilde g}_{(MAP)}\left[\phi^+_i{\tilde {\cal D}}_{\mu} \phi_i - 
\phi_i{\tilde {\cal D}}_{\mu}\phi^+_i\right],\lb{52}\ee 
where 
\be {\alpha}_{(MAP)}= \frac {\alpha}{N-1}\qquad {\mbox{and}}\qquad {\tilde\alpha}_{(MAP)} 
= \frac {\tilde\alpha}{N-1}.\lb{53}\ee 
The behaviour of the $SU(3)\times\widetilde{SU(3)}$ total beta--function, given by Figure 5 in 
the previous Section, is valid for the case of dyons.

{\it Dual symmetry is absent in nonabelian theories with matter fields.}

\section{Conclusions}

\begin{itemize}
\item[1.] In the present investigation we have suggested the Zwanziger type action for pure 
nonabelian theories.
\item[2.] We have shown that this action reveals the generalized dual symmetry of nonabelian 
theories and confirms the invariance under the interchange:
$$\alpha\to\tilde\alpha = \frac {1}{\alpha}.$$
\item[3.] Such a symmetry leads to the generalized renormalization group equations:
$$\frac{\mbox{d}\ln \alpha(t)}{\mbox{dt}} = - \frac{\mbox{d}\ln {\tilde \alpha}(t)}{\mbox{dt}}
= \beta(\alpha ) - \beta(\tilde \alpha) = \beta^{(total)}(\alpha)$$
with the total beta--function $ \beta^{(total)}(\alpha),$ which for pure nonabelian theories 
is {\it antisymmetric} under the interchange:
$$\alpha\leftrightarrow\tilde\alpha,\qquad {\mbox{or}}\qquad\alpha\leftrightarrow\frac 
1{\alpha}.$$
\item[4.] We have shown that, as a result of the dual symmetry, $\beta^{(total)}(\alpha)$ has 
zero at $\alpha = \tilde \alpha = 1$: 
$$\beta^{(total)}(\alpha=\tilde \alpha = 1) = 0.$$
\item[5.] We have applied the method of the Maximal Abelian Projections by t'Hooft to the 
description of the total beta--function in the case of the pure $SU(3)$ gauge theory, and 
demonstrated the behaviour of this beta--function in the region:
$$ 0 \le \alpha,\tilde\alpha < \infty.$$
\item[6.] At the first step we have considered the existence of the $N-1$ Higgs abelian scalar 
monopole fields ${\tilde \phi}_{1,2,..., N-1}$ and $N-1$ Higgs abelian scalar electric fields 
$\phi_{1,2,..., N-1}$ in the non--perturbative region of the pure nonabelian $SU(N)$ gauge 
theories.
\item[7.] At the second step we have assumed that the generalized dual symmetry naturally leads 
to the existence of the Higgs scalar dyon fields $\phi_{1,2,..., N-1}$, which are created in the
non--perturbative region of the pure $SU(N)$ gauge theories by non--perturbative effects of 
gluon fields. These abelian dyons describe the total beta--function in the non--perturbative 
region. For the pure $SU(3)$ gauge theory this non--perturbative region is:
$$ 0.4 \le\alpha\le 2.5,$$
what explains the freezing of $\alpha_s$ in QCD.
\end{itemize}

\section*{Acknowledgements}

The speaker is indebted to the Niels Bohr Institute (Copenhagen, Denmark) and to the Institute 
of Mathematical Sciences (Chennai, India) for their hospitality and financial support. This work
was supported by the Russian Foundation for Basic Research (RFBR), project No. 05--02--17642.

\end{document}